\documentclass[a4paper,fleqn,usenatbib]{mnras}

\usepackage{newtxtext,newtxmath}

\usepackage[T1]{fontenc}
\usepackage{ae,aecompl}

\usepackage{graphicx}	
\usepackage{amsmath}	

\title[The rotation rate of solar active and ephemeral regions]{The rotation rate of solar active and ephemeral regions -- I. Dependence on morphology and peak magnetic flux}

\author[A. S. Kutsenko]{
Alexander S. Kutsenko\thanks{E-mail: alex.s.kutsenko@gmail.com}
\\
Crimean Astrophysical Observatory, Nauchny, Crimea, 298409, Russia\\
}

\date{Accepted XXX. Received YYY; in original form ZZZ}

\pubyear{2020}

\begin{document}
\label{firstpage}
\pagerange{\pageref{firstpage}--\pageref{lastpage}}
\maketitle

\begin{abstract}
Using magnetic field maps acquired by the {\it Helioseismic and Magnetic Imager} on board the {\it Solar Dynamics Observatory} we measured rotation rates of 864 active and 322 ephemeral regions observed between 2010 and 2016. We found smaller magnetic tracers to show a tendency to rotate faster as compared to larger ones. Thus, ephemeral regions exhibit on average the fastest rotation rate. We further divided active regions into three classes. Class A comprised magnetic bipoles obeying Hale's polarity law, Joy's law, and exhibiting more coherent leading polarity in comparison with the following one. The second class B included active regions violating at least one of the aforementioned empirical laws. The third class U comprised unipolar active regions. We found no significant difference between the rotation rates of active regions of classes A and B. In contrast, unipolar active regions exhibited on average lower rotation rate and narrower distribution of the rotation rate differences. Assuming the rotation rate to indicate the anchoring depth of the magnetic structure within the convection zone, we supposed that active regions of classes A and B might be anchored throughout the entire convective envelope while unipolar active regions a rooted within a thin layer located either near the base of the convection zone or at a shallow near-surface depth.
\end{abstract}




\section{Introduction}
\label{sec:intro}

Solar differential rotation is an essential ``ingredient'' of the most of the global dynamo theories \citep[e.g.,][]{Charbonneau2014}. The differential rotation is presumably responsible for converting global poloidal magnetic field to the toroidal one (the so-called $\Omega$-effect). That is why a lot of attention is paid to precise measurements of the solar differential rotation since it may provide constraints for the solar dynamo models.

The rotation rate of the plasma within the convection zone varies with depth, latitude, and phase of the solar cycle \citep[e.g.][]{Pulkkinen1998, Li2013}. There are several approaches that are widely used to probe the solar differential rotation \citep{Beck2000}. The differential rotation can be estimated by measuring the Doppler shift of photospheric spectral lines. These measurements provide data on the angular velocity of the surface plasma \citep[e.g.][]{Howard1983}. Another approach is the tracking of any kind of tracers observed in the solar atmosphere. The tracers could be sunspots \citep[e.g.][]{Newton1951, Ward1966, Poljancic2017}, supergranules \citep[e.g.][]{Duvall1980, Snodgrass1990}, coronal bright points \citep{Sudar2015}, magnetic features of different sizes \citep[e.g.][]{Xu2016, Lamb2017}, coronal holes \citep[e.g.][]{Hiremath2013, Bagashvili2017} and other structures.

Finally, helioseismology is another way to probe the differential rotation of the Sun. As an advantage compared to other techniques, helioseismology provides data on the depth variations of the plasma angular velocity within the whole convective envelope \citep[e.g.][]{Thompson1996, Schou1998, Howe2000}. Although the exact details of the rotation profiles in the convection zone derived by different authors may differ, the qualitative picture of the differential rotation in the solar interior is quite similar. According to the helioseismology inversions, the radiative core rotates as a solid body while the differential rotation takes place above the tachocline layer from about 0.7~$R_{\sun}$ up to the surface. The rotation rate of the convection zone layers gradually increases for depths 0.8--0.95~$R_{\sun}$ within 0--45 degree latitudes reaching the fastest rotation rate at depths corresponding to 0.95--0.97~$R_{\sun}$. At the shallower depths of 0.97--1~$R_{\sun}$ an abrupt decrease of the rotation rate presumably takes place at all latitudes below 60 degrees \citep[see, e.g., fig.~1 in][]{Howe2000}.

The comparison of the rotation rates derived by feature tracking emphasizes different rotation rate values for different type of tracers. Since most of the tracers are related to the magnetic fields, the observed difference in the rotation rate might be attributed to different rates of the internal layers of the convection zone where these magnetic fields are anchored \citep[e.g.][]{Rhodes1990, Sivaraman2004}. Since sunspots were the first \citep[used by R.C.~Carrington back in 1863 as discussed in][]{Beck2000} and the most evident tracers to study solar differential rotation, the dependence of the sunspot rotation rates on their age, morphology, and size were analysed in details. 

\citet{Ward1966} used the Greenwich Heliographic Results to measure the rotation rate of sunspot groups observed between 1905 and 1954. He concluded that large groups tend to rotate slower than small ones. The rotation rate also depended on the morphology of a sunspot group. \citet{Ward1966} also showed that short-living non-recurrent sunspot groups rotate more rapidly than recurrent groups analysed by \citet{Newton1951}. \citet{Zappala1991} utilised the Greenwich Heliographic Results data covering the time interval between 1874 and 1976 and found young sunspot groups to rotate faster than recurrent ones. Similar conclusion were drawn in a number of other works \citep[e.g.][]{Balthasar1982, Ruzdjak2004}. \citet{Howard1992} found that individual sunspots exhibit stronger dependence of the rotation rate on the size as compared to sunspot groups. He also concluded that the rotation rate of sunspot groups depends on the polarity separation: groups with larger polarity separation tend to rotate faster. \citet{Howard1984} analysed 62-year long series of sunspot observations acquired at the Mount Wilson observatory. The authors found large sunspots to rotate slower than small sunspots. \citet{Howard1990} used Mount Wilson magnetic field data rather than white-light images to analyse the differential rotation of magnetic fields. He concluded that active regions rotate slower than faculae do. \citet{Howard1990} also found no significant difference between the rotation rates of leading and following parts of active regions. \citet{Pulkkinen1998}  carried out the most solid analysis of the sunspot rotation rates using the data from four databases covering the observational interval between 1853 and 1996. The authors found the rotation rate of sunspots to slowing down with time with no respect to the mean size of the groups. Summarizing the results obtained prior to 2000s, \citet{Beck2000} concluded that small sunspots rotate faster than large sunspots, young sunspots rotate faster than recurrent sunspots while supergranules rotate faster than most sunspots. \citet{Beck2000} assumed young sunspots to be anchored at a depth of about 0.93 $R_{\sun}$ while older sunspots disconnect from there roots and drift to shallower layers that rotate slower.

An opposite conclusion was drawn by \citet{Javaraiah1997} who analysed the rotation rate of sunspots in the Greenwich database observed between 1874 and 1976. They found that the rotation rate of large long-living sunspot groups tend to increase as the sunspot group ages. These results were interpreted as the rising of the anchoring layer of the long-living sunspot groups that were initially formed near the base of the convection zone. Short-living sunspot groups were found to rotate faster than long-living ones presumably due to anchoring near the surface. Similar results were obtained by \citet{Hiremath2002} who found long-lived sunspot groups (more than 6 days) to rotate faster from day to day. Acceleration of long-lived sunspot groups with age was further supported by \citet{Sivaraman2003} who utilised the Kodaikanal and Mount Wilson sunspot group data. The key difference of that study in comparison with previous works was life-span-wise sorting of sunspot groups. According to \citet{Sivaraman2003}, short-lived sunspot groups are anchored in the near-surface shallow layers.

A thorough analysis of the rotation rates may provide hints on the preferable depth where the most effective magnetic field generation -- the solar dynamo -- takes place \citep[e.g.][]{Brandenburg2005}. The global dynamo may operate near the tachocline layer, at the shallow near-surface leptocline layer or within the entire bulk of the convection zone. As an example, \citet{Nagovitsyn2018} found sunspot groups to form two subsets with statistically different differential rotation laws. This finding may be interpreted as the existence of two depths where these subsets of sunspot groups are formed. \citet{Brandenburg2005} considered faster rotation of young sunspot groups in comparison with older ones as an evidence in the favour of the near-surface global dynamo: young sunspots are presumably anchored at the fastest near-surface layers in the convection zone at a distance of about 0.95~$R_{\sun}$ from the Sun centre. As the sunspot group ages, the magnetic flux bundle forming the group rises to slower near-surface layers. In contrast, the generation of the magnetic flux bundles near the tachocline layers was argued by, e.g., \citet{Hiremath2002} and \citet{Sivaraman2003} basing on their rotation rates measurements. Note, however, that the anchoring of a sunspot at a certain depth does not necessarily require the magnetic flux bundle forming the sunspot to be generated at that depth \citep{Norton2013}. Moreover, all the above inferences assume that (i) the magnetic flux bundle forming an active regions emerges at the same latitude where it was generated in the convection zone and (ii) the roots of the magnetic flux bundle rise toward the surface as the active region ages.

Although the anchoring hypothesis if often used to interpret the difference in the rotation rates of magnetic tracers, this naive concept is not directly supported by theoretical deductions. On the contrary, numerical simulations of the magnetic flux loop emergence succeed in the explanations of the variations of the measured rotation rates without involving the anchoring assumptions, at least during the early stages of active region evolution. Visible rotation rates of young active regions are governed by a complex interplay of different forces and effects \citep{Petrovay2010}. The Coriolis force acting on the magnetic flux loop emerging through the convection zone causes different inclination of the following and leading legs of the loop, the former being more vertical while the latter being more inclined toward east \citep{Choudhuri1989, Fan1994, Caligari1995}. As a result of this geometrical asymmetry, during the emergence the leading part of the loop moves faster toward western direction from the point where the apex of the tube outbreaks. The eastward movement of the following part is more gradual \citep{vanDrielGesztelyi1990}. The apparent rotation rate of the whole active region appears accelerated due to proper motion of the legs. As the emergence of the loop proceeds, the expansion of the loop gradually reduces and the apparent rotation rate slows down. In addition, the westward-directed azimuthal component of the Coriolis force acting on the plasma downflows inside emerging flux tube ``results in a wavelike translational motion of the loop as a whole compared to the ambient medium: the active region is then expected to rotate faster than quiet sun plasma'' \citep{Petrovay2010}. Nevertheless, the anchoring may affect the rotation rate to some extent since the emerging magnetic flux bundles are anchored in the downdraft lanes of the giant convection cells up to dozens of Mm beneath the photosphere \citep{Chen2017}. Convective flows may also govern the rotation rate of magnetic tracers during late phases of their lifetime when a dynamical disconnection of the magnetic roots presumably takes place as suggested by \citet{Schussler2005}.

As one can see, there are still certain discrepancies in the interpretation of the rotation rate of sunspots groups of different age, size, and morphology. Moreover, the accuracy of the measurements of the sunspot groups rotation rate is also questioned \citep[e.g.][]{Petrovay1993}. These issues are worth to be addressed using present-day instruments. Although the time coverage of these instruments is sufficiently less than that offered by the sunspots databases, high cadence and high spatial resolution of modern uninterruptible observations allow accurate measurements of the differential rotation with unprecedented accuracy \citep[][]{Lamb2017}. Beside, as an advantage compared to continuum images, magnetic field maps allow tracking of small magnetic features and mature active regions simultaneously by the same routines. 

Our aim in this work is to measure the rotation rates of active and ephemeral regions observed during Solar Cycle 24 using magnetic field data acquired by the {\it Helioseismic and Magnetic Imager} \citep[HMI,][]{Schou2012} on board the {\it Solar Dynamics Observatory} \citep[SDO,][]{Pesnell2012}. We intend to explore the relationship between the rotation rate and the peak magnetic flux as well as between the rotation rate and morphology of active regions. We hope that our findings will provide additional constraints for dynamo models.

\section{Data and Methods}
\label{sec:data}

In this study we utilised data on line-of-sight (LOS) magnetic field acquired by SDO/HMI. SDO/HMI is a filter-type magnetograph that uses photospheric Fe~\textsc{i} 6173.3 \AA\ line to derive the vector magnetic field. Solar LOS magnetic field maps could be provided as full-disc 4096$\times$4096 pixel magnetograms with 720 s cadence. The spatial resolution of the instrument is 1 arcsec with the pixel size of 0.5$\times$0.5 arcsec$^{2}$. An automated feature-recognition algorithm identifies active regions on the full-disc magnetograms and assigns the so-called HMI active region patches \citep[HARP,][]{Bobra2014} numbers.

Active and ephemeral regions observed between 2010 May 05 and 2016 December 31 were selected as tracers for the rotation rate measurements. The list of tracers and some of their measured parameters are presented in Table~\ref{table0}. The full table is available as a supplementary material.

In this work we define ephemeral regions as small magnetic bipoles (emerging and evolving amidst quiet-Sun areas) with no assigned NOAA/HARP number.  In contrast, we assume active regions to be magnetic structures with assigned HARP number. Note that the weakest active regions with assigned HARP number not necessarily had assigned NOAA number.

Each tracer in the full-disc magnetogram was bounded manually by a rectangular box. By a cross-correlation technique the tracer was tracked on the consecutive (both back and forth in time) magnetograms as long as the centre of the bounding box was located within 60 degrees from the central meridian. All the tracers considered in the analysis were isolated in the sense that there were no crossing of the bounding box by another tracers or some significant portion of the magnetic flux during the entire time of observation. Note also that one HARP could contain several active regions. In this case each active region within one HARP was considered as a separate tracer and the HARP/NOAA numbers listed in the Table~\ref{table0} are for reference only.

\begin{table*}
	\centering
	\caption{The list of tracers and some of their measured parameters. Tracer label is a unique identifier used to distinguish the tracers during calculations. Tracer type are related to the Hale classification of active regions: 1 -- bipolar $\beta$ active region; 2 -- multipolar active region with $\gamma$ configuration; 3 -- multipolar active region with $\delta$ configuration; 4 -- unipolar $\alpha$ active region; 5 -- ephemeral region. The full table is available as supplementary material.}
	\label{table0}
	\begin{tabular}{lrrcrrccc} 
		\hline
		Tracer label&NOAA  &HARP  &Tracer&Peak  & Mean               & Synodical rot.  & Synodical rot.& Tracer sidereal\\
		 &number&number&type  &flux, & latitude,& rate of positive&rate of negative&rotation rate\\
		 &&&&10$^{21}$ Mx & deg & polarity, deg~day$^{-1}$&polarity, deg~day$^{-1}$&$\omega_{sid}$, deg~day$^{-1}$\\
		\hline
		20100504\_003600\_W04S26 & 11066 & 10 & 1 & 2.47 & -26.2 & 13.649$\pm$0.016 & 12.972$\pm$0.010 & 14.276$\pm$0.012\\   
		20100508\_004800\_E31S22 & 11071 & 17 & 1 & 0.92 & -21.2 & 13.596$\pm$0.009 & 13.281$\pm$0.009 & 14.401$\pm$0.008\\   
		20100504\_003600\_W37N24 & 11067 &  1 & 1 & 5.21 &  23.8 & 12.936$\pm$0.010 & 13.185$\pm$0.002 & 14.026$\pm$0.004\\   
		20100508\_004800\_W23S20 & 11068 & 12 & 1 & 5.66 & -19.8 & 13.104$\pm$0.003 & 13.441$\pm$0.005 & 14.236$\pm$0.003\\   
		20100523\_010000\_E37S05 &     0 &  0 & 5 & 0.20 &  -4.5 & 13.812$\pm$0.022 & 12.782$\pm$0.045 & 14.252$\pm$0.025\\   
		20100523\_010000\_W03S16 & 11072 & 26 & 1 & 7.18 & -15.1 & 13.574$\pm$0.002 & 13.629$\pm$0.004 & 14.555$\pm$0.001\\   
		20100527\_004800\_E17S29 &     0 &  0 & 5 & 0.17 & -28.4 & 12.631$\pm$0.023 & 12.901$\pm$0.026 & 13.719$\pm$0.015\\   
		20100530\_010000\_E20N15 &     0 &  0 & 5 & 0.28 &  14.7 & 13.146$\pm$0.018 & 11.922$\pm$0.025 & 13.486$\pm$0.018\\   
		20100530\_010000\_E04S19 & 11075 & 40 & 1 & 1.84 & -19.6 & 13.385$\pm$0.009 & 13.252$\pm$0.009 & 14.270$\pm$0.005\\   
		20100530\_010000\_W33N12 & 11073 & 38 & 1 & 3.11 &  12.2 & 13.452$\pm$0.005 & 13.423$\pm$0.006 & 14.389$\pm$0.001\\   
		\hline
	\end{tabular}
\end{table*}

To ensure the homogeneity of the data, ephemeral regions were manually cropped from the magnetograms in the same way as the active regions were processed. Although the number of ephemeral regions observed on the Sun significantly exceeds that of active regions, we selected a sample of only 322 ephemeral regions that is comparable to the sample size of active regions.

For each tracer, we calculated the temporal variation of the total unsigned magnetic flux by summing the absolute values of the magnetic flux density in pixels multiplied by a pixel area. Before calculating the magnetic flux, the maps of the magnetic field were preprocessed by applying a $\mu$-correction to each pixel as described in \citet{Leka2017}. The summing was performed over the pixels with the absolute magnetic flux density exceeding 18 Mx~cm\textsuperscript{-2} that is a threefold noise level in SDO/HMI LOS 720-s magnetograms \citep{Liu2012}.

\begin{figure}
	\includegraphics[width=\linewidth]{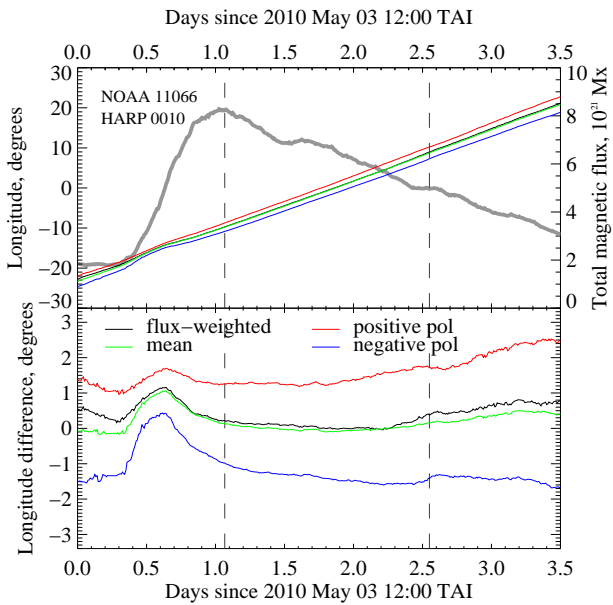}
    \caption{
    Top -- The total unsigned magnetic flux (thick gray curve) evolution of NOAA active region 11066. Coloured lines show the longitudinal positions of flux-weighted positive (red) and negative (blue) polarities of the active region. Mean unweighted longitude and flux-weighted longitude of the active region are shown in green and black, respectively. Bottom -- The difference between the longitudes of the tracer and some point rotating at a constant rate. The colour coding is the same as in the top panel. Vertical dashed lines in both panels denote the time interval within which the fitting of the $\theta(t)$ curve by a linear approximation was performed to measure the rotation rate.
    } 
    \label{fig0}
\end{figure}

To measure the rotation rate we applied the following procedure. In order to decrease the uncertainties due to noise in the magnetic field maps, the linear size of the cropped patches was reduced by a factor of two by binning 2$\times$2 pixels. Then the flux-weighted centroid positions of each magnetic polarity were calculated. To decrease the uncertainties only pixels with absolute values exceeding 100 Mx~cm\textsuperscript{-2} were used in the calculation. The positions of the centroids were further converted to Stonyhurst heliographic coordinates using World Coordinate System (WCS) library provided in the \textsc{IDL SolarSoft} package. The longitude of the tracer, $\theta(t)$, was determined as the unweighted mean value of the longitudes of positive and negative polarities. This is the difference between our methodology and procedures widely applied in the majority of the previous studies where the flux-weighted or area-weighted centre of the tracer was considered with no respect to leading and following parts. \citet{Petrovay1993} argued that due to different decay rate of leading and following polarities the area-weighted or flux-weighted centre of an active region shifts toward stronger leading polarity as the active region evolves \citep[see fig.~2 in][]{Petrovay1993}. This effect may cause fake proper motion of the tracer yielding polluted rotation rate measurements. An example is shown in Fig.~\ref{fig0}. The total unsigned magnetic flux of NOAA active region 11066 is plotted by a thick grey line in the top panel of Fig.~\ref{fig0} while the flux-weighted longitudes of negative and positive magnetic polarities are shown by blue and red lines, respectively. Green line denotes unweighted mean longitude of the active region while the flux-weighted longitude is shown in black. The differences between the mentioned longitudinal positions and some point rotating at a constant rate are shown in the bottom panel of Fig.~\ref{fig0}. One can see that, as expected, negative and positive polarities disperse from each other as the active region evolves. Meanwhile the flux-weighted position exhibits motion toward the leading positive-polarity part of the active region as decaying proceeds.

The synodical rotation rate, $\omega_{syn}$, of the tracers was measured by fitting the unweighted longitude, $\theta(t)$, by a linear approximation. In order to decrease the influence of the proper motion of magnetic polarities during the active and ephemeral regions emergence (described in Section~\ref{sec:intro}), the fitting was performed within the time interval when the tracers were in a ``mature'' state (shown by dashed vertical lines in Fig.~\ref{fig0}). Thus, where possible, we discarded the emergence phase and used the time interval from the observed peak magnetic flux until the magnetic flux decrease to approximately a half of the peak value. Consequently, the rotation rate measured in this work should be considered as the mean rotation rate averaged over the interval of the maximum development of an active or ephemeral region. Since the following magnetic polarity of unipolar regions is usually very weak and dispersed, the measurement of its flux-weighted centre is unreliable. Therefore, the rotation rate of unipolar tracers was measured by fitting the $\theta(t)$ curve of the magnetic polarity (either negative or positive) coinciding with the polarity of the unipolar region itself. The peak total unsigned magnetic flux, $\Phi_{max}$, observed within the fitting interval was stored for further analysis.

Following \citet{Lamb2017}, we applied the procedure proposed in \citet{Skokic2014} to calculate the tracer sidereal rotation rate
\begin{equation}
    \omega_{sid} = \omega_{syn} + \frac{\omega_{Earth}}{r^2} \frac{\cos^2 \Psi}{\cos i},
\end{equation}
where $\omega_{Earth}$ is the Earth's yearly averaged angular velocity, $r$ in the dimensionless distance between the Sun and Earth in a.u., $\Psi$ is the angle between the pole of the ecliptic and the solar rotation axis orthographically projected onto the solar disc \citep{Skokic2014}. Calculation of $\Psi$ requires the knowledge of ephemerides that were retrieved from the JPL Horizons website at \verb|https://ssd.jpl.nasa.gov/horizons.cgi|. See \citet{Skokic2014} and \citet{Lamb2017} for more details on the $\omega_{sid}$ calculations.

\section{Results}

In all, the rotation rates of 864 active and of 322 ephemeral regions were analysed in this work. The peak magnetic flux varied between 0.3$\times 10 ^{21}$ Mx and 1.24$\times 10^{23}$ Mx for active regions and between 0.5$\times 10^{20}$ Mx and 4.8$\times 10^{21}$ Mx for ephemeral regions. The tracers were observed within the latitudinal ranges of approximately $\pm$40 degrees.

\begin{figure*}
	\includegraphics[width=\linewidth]{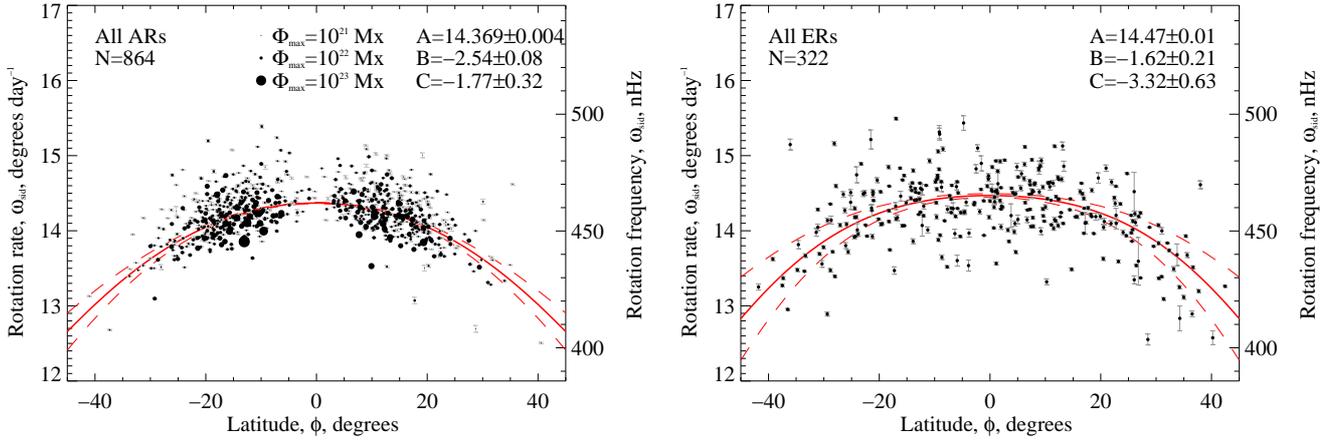}
    \caption{Sidereal rotation rate of active (left panel) and ephemeral (right panel) regions versus heliographic latitude. The size of the circles in the left panel is proportional to the peak total magnetic flux of active regions. Error bars show 1-$\sigma$ uncertainty of the linear fitting used to derive the rotation rate (see text). Red solid curves are the best non-linear fittings of the distributions using equation \ref{eq2}. Dashed red curves show 2-$\sigma$ uncertainties of the fittings. Differential rotation constants from equation \ref{eq2} and their uncertainties are listed in each panel.}
    \label{fig1}
\end{figure*}

Fig.~\ref{fig1} shows the measured tracer rotation rates versus heliographic latitude (black circles). The error bars are 1-$\sigma$ uncertainties of the $\theta(t)$ curve linear fitting, $\sigma_{\omega}$. We used the Levenberg-Marquardt algorithm to fit the distribution by a commonly-used function (red solid curves in Fig.~\ref{fig1})
\begin{equation}
    \omega_{sid} = A + B \sin^2 \phi + C \sin^4 \phi,
    \label{eq2}
\end{equation}
where $A$ is the equatorial rotation rate, $B$ and $C$ are responsible for the differential rotation. The weights of the data points were set to be 1/$\sigma_{\omega}$. The fitting yields the following differential rotation law for all active regions
\begin{equation}
    \omega_{sid} = 14.369 - 2.54\sin^2 \phi -1.77 \sin^4 \phi.
    \label{eq3}
\end{equation}

The visual comparison of the $\omega_{sid}$ versus $\phi$ distributions for active (left panel in Fig.~\ref{fig1}) and ephemeral (right panel in Fig.~\ref{fig1}) regions reveals, as expected, higher scatter of the ephemeral regions rotation rates. Seemingly, it can be explained by the governing role of the turbulent surface plasma flows that drag small magnetic tubes forming ephemeral regions. Nevertheless, on average, ephemeral regions rotate faster as compared to active regions. The equatorial rotation rate for ephemeral regions $A=14.47 \pm 0.01$ degrees~day\textsuperscript{-1} is higher than that for active regions, $A=14.369 \pm 0.004$ degrees~day\textsuperscript{-1}.

\begin{figure*}
	\includegraphics[width=\linewidth]{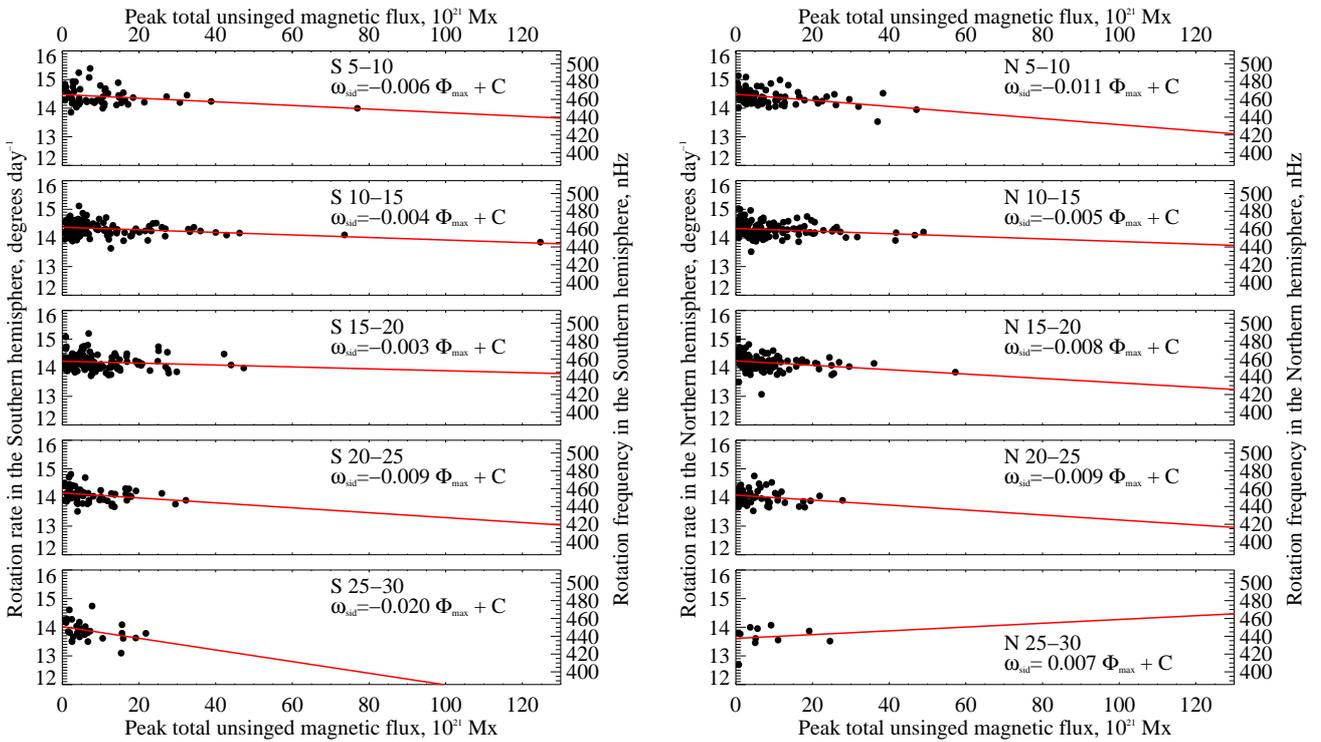}
    \caption{Sidereal rotation rate versus peak magnetic flux for active regions observed within different 5-degrees wide latitudinal belts in the Southern (left panels) and in the Northern (right panels) hemispheres. Red solid lines show the best linear fittings of the distributions. The slopes of the fittings are shown in each panel.}
    \label{fig2}
\end{figure*}

The area of the circles in the left panel of Fig.~\ref{fig1} is proportional to the peak unsigned magnetic flux of the tracers. Visual inspection of the figure hints at higher rotation rate for smaller active regions. To confirm this inference we divided all the active regions into 5-degrees wide latitudinal belts and explored the relationship between the rotation rate and peak unsigned magnetic flux within each belt. The results are presented in Fig.~\ref{fig2}. Indeed, although the tendency is very weak (the absolute value of Pearson's $R$ does not exceed 0.25 for each latitudinal belt) linear fittings of the distributions indicate that stronger active regions tend to rotate slower. The only exception is the 25--30 degrees latitude belt in the Northern hemisphere, however the data are scanty within these latitudinal ranges.

We further analysed the dependence of the rotation rate on the morphology of active regions. All active regions with assigned NOAA numbers were divided into three classes introduced in \citet{Abramenko2018}. The first class A comprises active regions obeying three empirical laws: (i) magnetic polarity of leading and following parts are in accordance with the Hale's law; (ii) tilt of the magnetic bipole axis is in an agreement with Joy's law; and (iii) the leading polarity is more coherent as compared to the following one. The existence of such active regions is explained by a phenomenological Babcock-Leighton mechanism \citep{Babcock1961, Leighton1969} or mean-field \citep{Steenbeck1969} global dynamo theory. Recall that these models assume initial magnetic fields to be generated at the base of the convection zone. The second class B includes active regions violating at least one of the aforementioned empirical laws. We hypothesize that the appearance of active regions of class B may be related to local fluctuations or  disturbances of the magnetic flux bundle by a turbulent plasma during its generation or buoyant emergence. Finally, the last class U consists of unipolar sunspots. Only unipolar active regions exhibiting neither pores nor spots in the following part in the white-light images were added to class U subset. The following polarity of class U active regions in magnetograms was just a disperse supergranular network during the entire interval of observation. By dividing active regions into these three classes, we intend to reveal whether there exist some preferential depth where the certain type of active regions is formed and anchored. The comparison of the anchoring depths may provide clues for models of the dynamo that is responsible for generation of these magnetic structures.

\begin{figure}
	\includegraphics[width=\columnwidth]{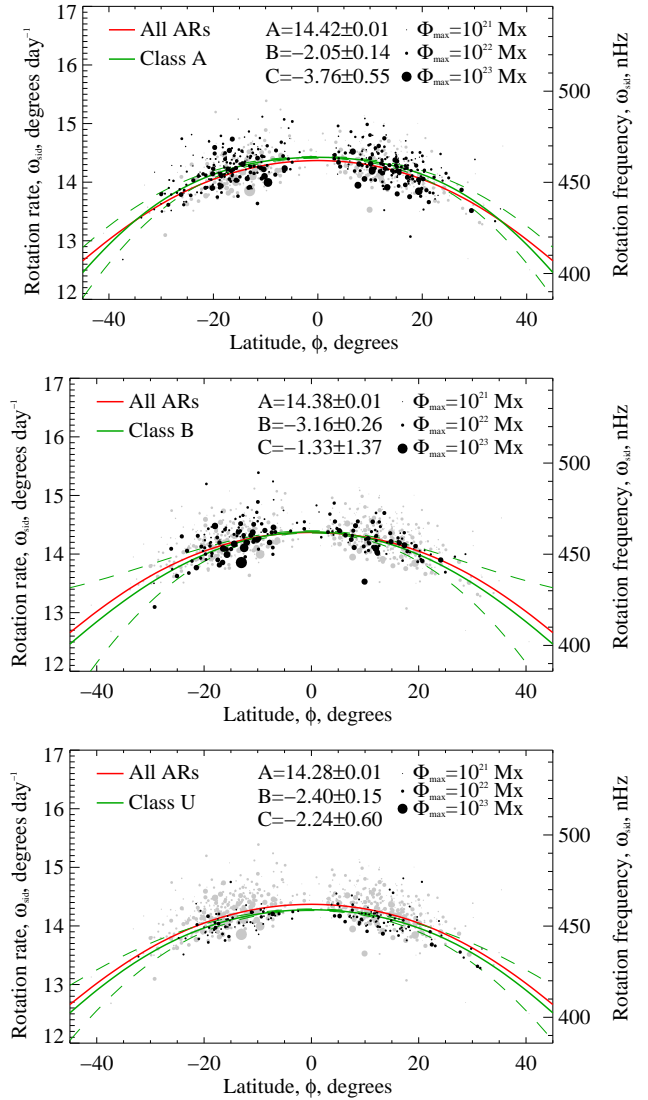}
    \caption{Sidereal rotation rate versus heliographic latitude for active regions of classes A (top), B (middle), and U (bottom). The area of the circle is proportional to the peak total magnetic flux of an active region. All active regions are denoted by gray circle while active regions of a certain class are shown in black. Green solid curves are the best non-linear fittings of the distributions using equation \ref{eq2} of a certain class of active regions. Dashed green curves show 2-$\sigma$ uncertainties of the fittings. Differential rotation constants from equation \ref{eq2} and their uncertainties are listed in each panel. Red curve in each panel shows the rotation rate derived from the fitting of all active regions (equation \ref{eq3}) for comparison.}
    \label{fig3}
\end{figure}

Fig.~\ref{fig3} shows the rotation rates for active regions of classes A (top panel), B (middle panel), and U (bottom panel). Green curves are non-linear fittings (expression \ref{eq2}) of the active region rotation rates. The fitting of all active regions is shown in red for comparison. Fig.~\ref{fig3} suggests that there is no obvious difference between the rotation rates of active regions of classes A and B. Meanwhile, unipolar active regions lie predominantly below the red curve in the bottom panel of Fig.~\ref{fig3} suggesting lower rotation rates as compared to other active regions. In addition, the size of symbols of unipolar active regions indicates also lower peak magnetic flux in comparison with active region of classes A and B.

\begin{figure*}
	\includegraphics[width=\linewidth]{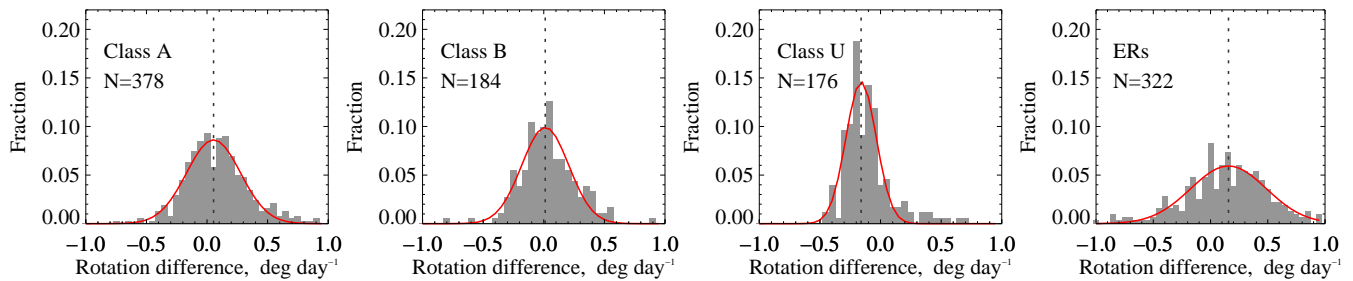}
    \caption{Normalised distributions of the rotation difference (see text) derived for (from left to right) active regions of class A, class B, class U, and for ephemeral regions. The number of tracers used to create the histogram is shown in each panel. Red solid curve shows the best Gaussian fitting of the histogram in each panel. Dashed vertical lines denote the modes of the distributions.}
    \label{fig4}
\end{figure*}

Fig.~\ref{fig4} demonstrates the normalised distribution of the rotation difference for each subset of tracers considered in this work. To obtain the rotation difference, the mean rotation rate calculated using equation \ref{eq3} was subtracted from the rotation rate of each individual tracer. Active regions of classes A and U demonstrate Gaussian-like distributions. In contrast, the rotation difference distribution of class B active regions is more intermittent and irregular. The reasons of these irregularity requires further studies.

\begin{figure*}
	\includegraphics[width=\linewidth]{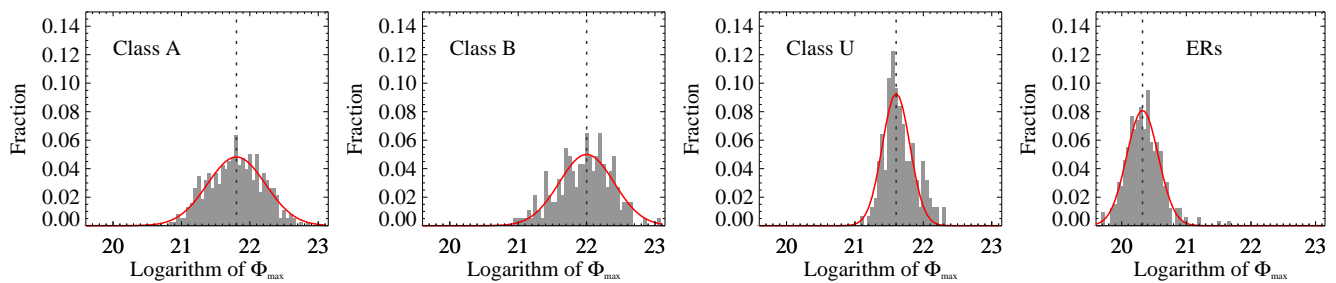}
    \caption{Normalised distributions of the decimal logarithm of the peak magnetic flux observed for (from left to right) active regions of class A, class B, class U, and for ephemeral regions. Red solid curve shows the best Gaussian fitting of the histogram in each panel. Dashed vertical lines denote the modes of the distributions.}
    \label{fig5}
\end{figure*}

Similar normalised distributions of the peak magnetic flux for active regions of different classes as well as for ephemeral regions are shown in Fig.~\ref{fig5}. Active regions of class B exhibit on average the highest peak magnetic flux suggesting that the disturbances of a regular magnetic flux bundle (presumably responsible for the formation of active regions of class A) might be associated with the magnetic flux growth.

Each distribution in Figs.~\ref{fig4} and \ref{fig5} was approximated by a Gaussian function. Peak values of these approximations (modes of the distributions) as well as rotation rate constants for each subset of tracers are summarized in Table~\ref{table1}. The widths of the Gaussians plotted in Fig.~\ref{fig4} are also listed in the 7th column of Table~\ref{table1}.

Comparison of the histograms in Fig.~\ref{fig4} and of their parameters (columns 6 and 7 in Table~\ref{table1}) suggests that there is no significant discrepancy between the rotation rate distributions of active regions of classes A and B. Most of active regions of these classes exhibit near-zero rotation difference. The widths of the rotation difference distributions are also quite similar. Unipolar active regions (class U) form narrower rotation difference distribution as compared to bipole and multipole active regions. In addition, the shift of the class U histogram in Fig.~\ref{fig4} towards negative values suggest that most of unipolar active regions exhibit slower rotation rate. Meanwhile, ephemeral regions exhibit the fastest rotation and the widest scatter of the rotation rates.

Average peak magnetic flux of active regions decreases from class B to class A to class U (column 8 in Table~\ref{table1}). Interestingly, unipolar active regions (class U) exhibit, on average, the lowest peak magnetic flux and the lowest rotation rate that is in a contradiction with the tendency shown in Fig.~\ref{fig2}. This issue is addressed in more detail in the next section.

\begin{table*}
	\centering
	\caption{Derived differential rotation constants and other parameters of the tracers analysed in this work.}
	\label{table1}
	\begin{tabular}{lccccccc} 
		\hline
		Tracer & Number &A & B & C & Rotation difference& Rotation difference& Peak magnetic\\
		type & of tracers &deg day$^{-1}$ & deg day$^{-1}$ & deg day$^{-1}$ & mode, deg day$^{-1}$ & FWHM, deg day$^{-1}$ & flux mode, 10$^{21}$ Mx\\
		\hline
		All active regions& 864 & $14.369 \pm 0.004$ & $-2.54 \pm 0.08$ & $-1.77 \pm 0.32$ & -- & -- & --\\
		Class A           & 378 & $14.42 \pm 0.01$ & $-2.05 \pm 0.14$ & $-3.76 \pm 0.55$ & 0.05 & 0.22 & 6.3\\
		Class B            & 184 & $14.38 \pm 0.01$ & $-3.16 \pm 0.26$ & $-1.33 \pm 1.37$ & 0.01 & 0.19 & 9.6\\
		Class U            & 176 & $14.28 \pm 0.01$ & $-2.40 \pm 0.15$ & $-2.24 \pm 0.60$ & -0.16 & 0.13 & 4.0\\
		Ephemeral regions  & 322 & $14.47 \pm 0.01$ & $-1.62 \pm 0.21$ & $-3.32 \pm 0.63$ & 0.16 & 0.33 & 0.2\\
		\hline
	\end{tabular}
\end{table*}

\section{Conclusions and Discussion}

We used LOS magnetic field maps acquired by SDO/HMI to measure the rotation rates of 864 active and of 322 ephemeral regions observed between 2010 and 2016. Each tracer was tracked at full-disc magnetograms by a cross-correlation technique. The peak magnetic flux of tracers varied between 0.3$\times 10^{20}$ and 1.24$\times 10^{23}$ Mx. We divided active regions into three classes A, B, and U. Recall that class A comprised magnetic bipoles obeying both Hale's law and Joy's law and exhibiting more coherent leading polarity as compared to the following one. Class B included magnetic bipoles and multipoles violated at least one of these empirical laws. Class U comprised unipolar active regions.

Our analysis of the rotation rates of different tracers yielded the following conclusions:
\begin{enumerate}
    \item
    Active regions exhibit sufficiently slower rotation as compared to ephemeral regions. Moreover, there exist a weak tendency for larger active regions to rotate slower. This finding is in agreement with previous conclusions made in other studies \citep[e.g.][]{Ward1966, Beck2000}.

    \item 
    Ephemeral regions exhibit higher scatter of the rotation rates as compared to active regions.

    \item
    Most of unipolar active regions rotate slower than the mean rotation rate at all latitudes analysed in this work. At the same time unipolar active regions exhibit on average lower peak magnetic flux as compared to multipolar magnetic structures \citep[see also ][]{Ruzdjak2005}. Consequently, they disobey the rule found for all active regions regarding the tendency for weaker active region to rotate faster. 
    
    \item
    We found no significant difference between the parameters of the rotation rate distributions of active regions of classes A and B. These distributions exhibit similar widths and near-zero modes (Fig.~\ref{fig4} and columns 6, 7 in Table~\ref{table1}). In contrast, the rotation differences of unipolar active regions form relatively narrow distribution shifted toward negative values.
    
\end{enumerate}

We further discuss our results assuming the tracer rotation rate to be determined by the angular velocity of the anchoring layer. The depth of the anchoring layer can be estimated using the rotational profiles of the internal plasma as derived by helioseismology. By projecting the rotation rate of a tracer on the rotation profile of the internal plasma at a given latitude \citep[e.g. fig.~1 in][]{Howe2000} we can guess at which depth the anchoring layer is located. This issue was addressed in detail in \citet{Sivaraman2004}. 

We found ephemeral regions to exhibit relatively high scatter of the rotation rates. The reason for this might be the governing role of the photospheric plasma flows on the proper motions of weak magnetic structures. Nevertheless, on average, ephemeral regions rotate faster in comparison with larger active regions. The equatorial rotation rate for ephemeral regions is 14.47$\pm$0.01 degrees~day$^{-1}$ (the rotation frequency is approximately 465 nHz). This rotation rate is the highest observed within the convection zone and it corresponds to the depth of the leptocline layer at about 0.95~$R_{\sun}$ \citep[e.g.][]{Howe2000}. Note also that this rotation rate is higher than that listed by \citet{Lamb2017} in his table~1. Thus, \citet{Lamb2017} found the rotation rate to increase with for tracers with longer lifetime. The tracers with the lifetime of about 1 day (100 frames with 720 s cadence) in \citet{Lamb2017} exhibited mean rotation rate of 14.11 degrees~day$^{-1}$ within 20--22 degrees latitudinal belt. For the same latitudes our fitting of ephemeral regions yields 14.23 degrees~day$^{-1}$. The mean lifetime of ephemeral regions in this work is of about 2 days. Consequently, our results and that obtained in \citet{Lamb2017} are in a good agreement: weaker magnetic structures are presumably anchored closer to the solar surface. We suppose that magnetic bipoles with a lifetime of about several days and peak magnetic flux of about $10^{21}$ Mx are generated and initially anchored near the leptocline. Smaller magnetic structures are generated within shallower near-surface layers.

The similarity of the parameters of the rotation rate distributions of active regions of classes A and B might be explained by the same formation depths. The rotation frequencies vary in a wide range from 410 to 480 nHz suggesting that these active regions might be anchored throughout the entire convective envelope. Consequently, the distortions of the magnetic flux bundles, which are considered as reasons for the existence of class B active regions, might take place at all depths within the convection zone.

The most interesting result in our opinion is the rotation rates of unipolar active regions (class U). Narrower distribution of the rotation differences of unipolar active regions suggests that most of them are anchored within relatively thin layer in the convection zone. The equatorial rotation rate for tracers of class U is 14.28$\pm$0.01 degees~day$^{-1}$ (the rotation frequency is 459 nHz). This rotation rate corresponds to the anchoring layer located either at a distance of 0.75~$R_{\sun}$ or 0.98~$R_{\sun}$ from the Sun centre \citep[see fig.~1 in][]{Howe2000}. Assuming anchoring layer to rise with time one would expect these tracers to be anchored in the near-surface layers. However, to confirm this statement we should analyse the temporal variations of the rotation rates that will be the topic of our second paper.

\section*{Acknowledgements}

I would like to express my deep appreciation to Dr. Kristof Petrovay whose comments stimulated me to reconsider the methodology and improve the manuscript. I am also grateful to Dr. V.I.~Abramenko for fruitful discussions and careful reading of the manuscript, to Dr. A.V.~Zhukova for providing the classification of active regions. The HMI data are courtesy of NASA/SDO and the HMI Science Team. The study was supported by the Russian Science Foundation, Project 18-12-00131.

\section*{Data availability}

The data underlying this article are available in the article and in its online supplementary material.












\bsp	
\label{lastpage}
\end{document}